\newcommand{\beqn}{\begin{equation}}
\newcommand{\eeqn}{\end{equation}}
\newcommand{\beqa}{\begin{eqnarray}}
\newcommand{\eeqa}{\end{eqnarray}}

\input{german.sty}
\documentstyle[11pt]{article}
\let\eps=\epsilon
\let\al =\alpha
\let\sig=\sigma
\let\bet=\beta
\let\0=\"

\begin{document}
\title{Numerical Studies on the Magnetism of Fe-Ni-Mn
Alloys in the Invar Region
\thanks{based on the Diploma thesis of A.~Paintner,
Regensburg 1994}
}
 \author{A.~PAINTNER, F.~S"USS and U. KREY
\\ \\ Institut f\"ur Physik II der Universit\"at Regensburg,\\
Universit\"atsstr.~31, D-93040 F.R.G. }

%\begin{document}

\large

\maketitle

\begin{abstract} By means of self-consistent semi-empirical LCAO
calculations we study the itinerant magnetism of %face-centred
%%disordered\newline
(Fe$_{0.65}$Ni$_{0.35}$)$_{1-y}$Mn$_y$ alloys for $y$ between 0 and 0.22
 at $T=0$ K, neg\-lecting only the
transverse spin components. We find that the magnetic behaviour
is quite complicated on a local scale. In addition to ferromagnetic
 behaviour, also metastable spin-glass-like configurations are found.
In the same approach, using
 a direct numerical calculation by the Kubo-Formalism without any fit
parameters, we also calculate the electrical conductance in the magnetic state
and find that the $y$-dependence observed in the experiments is well
reproduced by our calculations, except of an overall factor of $\approx$ 5, by
which our resistivities are too large. %The discrepancy is perhaps due to
%a neglection of transport channels which involve transverse spin
%components.
\end{abstract}

\section{Introduction}

In two recent papers, \cite{Fruechtl,Suess}, we have reported progress in the
calculation of the itinerant magnetism of ordered or disordered metallic
alloys in a self-consistent semi-empirical LCAO approach ( LCAO= ''Linear
Combination of Atomic Orbitals''). In the first of
those papers, \cite{Fruechtl}, fcc iron and particularly disordered
fcc Fe$_{1-x}$Ni$_x$ alloys have been treated, with particular emphasis
on the so-called Invar region around $x=0.35$, where the local Fe moments
respond quite sensitively on changes of the local neighbourhood and on
the local volume. In fact, the Invar behaviour is essentially a
consequence of strong magneto-volume effects (see Wassermann,
\cite{Wassermann}, for a review), which happens both in face-centered  and also
in amorphous Fe, see \cite{Fruechtl} and \cite{Krauss}, respectively.

In the second paper, \cite{Suess}, we concentrated on elemental Mn and various
 Mn
alloys with Ni or Fe, respectively. Here it turned out that the magnetism
of Mn is very complicated, e.g.~$\al$-Mn has a primitive cubic elementary cell
with 58 atoms and a very complicated antiferromagnetic spin
configuration. This spin configuration is not even exactly known, but seems to
be
non-collinear with at least eight non-equivalent magnetic sites in the
elementary cell, where not only the Mn moment directions, but also the
magnitudes of the moments are strongly different for these sites (see
\cite{Suess} for details). However, in spite of the difficulties, we
finally obtained in \cite{Suess} a consistent description of the
parameters involved in our semi-empirical model. Here we used 9 orbitals
per atom, namely five 3d-, three 4p- and one 4s-orbitals,
and parametrized the hopping
 elements appearing in the Hamiltonian according to a Slater-Koster
 approach with Two-Center-Integrals up to third-nearest neighbours as
 given in the book of Papaconstantopoulos, \cite{Papacon}.
Furthermore, in our semi-empirical LCAO approach the
 atomic charges and moments are obtained  selfconsistently
 from a single intra-d-orbital Hubbard integral $U$, see
 \cite{Fruechtl,Suess,Krauss}. That is, the mechanism leading to the itinerant
 magnetism is the interplay of Coulomb interaction and Pauli principle,
 according to which only two electrons with different spins can occupy the
 same orbital, but if they do, this costs an additional Coulomb repulsion $U$,
 so that energy can be gained by having different occupation numbers in the up-
 and down spin states, respectively.

  At the end of paper \cite{Suess}, we finally also obtained consistency of
  the band shift parameters for the different kinds of atoms in binary alloys
of
  Mn, Ni and Fe, respectively.

    Thus we are now able to treat also the disordered ternary alloy
     (Fe$_{1-x}$Ni$_x$)$_{1-y}$Mn$_y$. Here we concentrate on the case
$x=0.35$,
     where without Mn one is in the above-mentioned Invar region, and where
     according to the experimentalists, with increasing Mn concentration $y$ a
     transition  to spin glass behaviour should happen
     (see \cite{Boettger,Boettgeretal,HesseBoett,Fricke}). The problem of this
     transition is of special concern to us in this paper.

     With our LCAO-formalism, we  have also been able to calculate
     directly the resistivity of the alloys in the magnetic state, which
     has been measured by the same experimental group,
     \cite{BarnardBoett}.

Our paper is organized as follows: In chapter 2, we describe our
formalism and details of the computation. In chapter 3,
respectively, the results for
the magnetic behaviour and for the resistivity are presented, and
finally in chapter 4, we give our conclusion and discuss remaining
open questions.

\section{Formalism}

Our alloys are modelled by 108 atoms, arranged on the sites of an fcc cube
with periodic boundary conditions. The occupation of the sites with Fe,
Ni and Mn atoms is performed randomly according to the formula
(Fe$_{1-x}$Ni$_x$)$_{1-y}$Mn$_y$. For every pair  of concentrations $x$
and $y$ considered in the
results below, typically ten different random samples have been generated, from
which averages have been gained. Concerning the lattice constants, we
took the results presented on p.~91 of \cite{Boettger}, according to
which the $y$-dependence of the lattice constant $a$ can roughly be
 described by a parabola passing through $a$=3.59
$\AA$ at $y=0$ and $y=0.19$, with the minimum value of
 $a=$ 3.584 $\AA$ at $y=0.095$
in-between.

 The quantum-mechanical self-consistency equations, which have to be
 solved in our LCAO approximation, are
 \beqn \label{eqn1}
 \sum_{m,\beta} H_{l,m}^{\al ,\beta} c_{m,\beta ,\sigma}^{(\nu )}+
U_{l,\al}\cdot \left (
 n_{l,\al ,-\sigma} -n_{l,\al}^{(para)} - \sig\cdot\Delta_{l} \right
 )\cdot
 c_{l,\al ,\sig}^{(\nu)} = \epsilon_{\sig}^{(\nu )} c_{l,\al ,\sig}^{(\nu)}\,.
 \eeqn

 Here  $l$ and $m$ count the 108 sites, $\al$ and $\bet$ the 9 orbitals
  per site, and  $\sig = \pm 1$ the two possible orientations of the spin;
 for every spin direction, the index $\nu$ counts the 9$\times $108 possible
 orthonormal single-particle eigenvectors $\vec{c}_{\sig}^{\,\,(\nu)}$
  with real components $c_{l,\al ,\sig}^{(\nu )}$ and
 the corresponding single-particle energies $\epsilon_{\sig}^{(\nu)}$.
 The parameter $U_{l,\al}$ is the above-mentioned Hubbard-type
 intra-orbital   Coulomb integral (see below), while $n_{l,\al ,\sig}$ and
 $n_{l,\al}^{(para)}$ are the expectation values of the occupation numbers in
 the magnetic and paramagnetic state, respectively, and determined by
 the self-consistency equations

\beqn \label{eqn2}
n_{l,\al,\sig} = \sum_{\nu =1}^{\nu_f(\sig )} |c_{l,\al ,\sig}^{(\nu)} |^2 \,.
\eeqn

 In eqn.~(\ref{eqn2}), the sum $\nu =1,...,\nu_f(\sig)$ is over
  the occupied single-particle
energies of the given spin-direction. In the present disordered magnetic
system ,
these states are non-degenerate.

Of course it should be mentioned that this treatment of the Hubbard
 interaction corresponds to a mean-field approximation, so that in
 the calculation of the total energy one would have to subtract the
well-known double counting correction terms from the sum of the
 single particle energies (see e.g.~\cite{Suess,Fruechtl}); however, as
already mentioned above, the energies calculated from our approach are
in any case not very reliable.

Finally, there appear parameters $\Delta_{l}$ in eqn.~(\ref{eqn1}),
which play the role of effective fields serving for the initialization
of the magnetic state, if one starts from a paramagnetic state. These
initializing fields are switched off after the first few iterations;
their position dependence becomes an important point, when one wants
to generate spin glass configurations (see below).

Concerning the values of $U_{l,\al}$, we assume as usual that they are
 different from 0 only for $d$-states and actually depend only on the
kind of atom considered, namely $U_{l,\al}$= 5.3 eV for Fe and Ni, see
 \cite{Fruechtl}, and \hbox{$2.9$ eV} for Mn, see \cite{Suess}.

          The $H_{l,m}^{\al ,\bet}$-matrix in eqn.~(\ref{eqn1})
          describes the LCAO Hamiltonian for the {\it paramagnetic} system.
Here, the elements with $l=m$, but $\al\ne\bet$, vanish, while for the
terms with $\al=\bet$, i. e. the self-energies,
we use the numbers given in
 %table I. These values are from
 the book of Papaconstantopoulos, \cite{Papacon} p. 289 ff. And as
already mentioned in \cite{Suess}, although the Fe values are taken
from bcc Fe, they work even for fcc alloys of Fe, Ni and/or Mn, if one
 introduces additional
self-energy
-shifts $d_{Fe\to  Mn}=+0.9$ eV and
 $d_{Fe\to Ni}=+0.95$ eV for the $H^{\alpha \alpha}_{l l}$-levels of Mn
resp.~Ni relative
 to the Fe levels. Of course, finally in the magnetic state the
 $n_{l, \alpha, \sigma}$ are in any case self-consistently
 determined through the $U$-terms in eqn.~(\ref{eqn1}).

 Concerning the hopping matrix elements $H_{l,m}^{\al ,\bet}$, i.e.~with
 $l\ne m$, we have again relied on the parametrization with two-center
 integrals and orthonormal orbitals in \cite{Papacon}, and again, as in
 \cite{Fruechtl,Suess}, we
 take into account the fact that the distances between the sites
  in our alloys are different  from those in \cite{Papacon} by modifying
 the two-center integrals $I_{\al,\bet;\,\gamma}$ (e.g.~with $\gamma =\sigma$,
 $\pi$ or $\delta$)
  according to the {\it ansatz}
  \beqn
       I_{\al ,\bet ;\,\gamma}(r)= \eta_{\al ,\beta ;\,\gamma}\cdot
 r^{-k} \,.
  \eeqn
  Here $k=5$ for ($\al ,\beta$)= (d,d); $k=7/2$ for ($\al ,\beta$)=
  (p,d) or (s,d), and $k=2$ for ($\al ,\beta$)=
  (s,s), (s,p) or (p,p); see Harrison \cite{Harrison}. The $\eta$-parameters
   are fitted from the $I$-values given
   in \cite{Papacon} for the distances there; they are different for the
   first, second and third neighbour shell. For  Fe, the situation
    is somewhat more complicated: There, the $\eta$-values for the
     three neighbour shells in fcc Fe are obtained from those contained
      in the results for bcc Fe  in \cite{Papacon} with the interpolation
       procedure given in the appendix of \cite{Kromp1}.

 This determines finally the paramagnetic part of our Hamiltonian:
 For the magnetic properties, in which we are finally interested,  and  also
 for the charge distribution in the magnetic state, self-consistency is
 then demanded, expressed by the terms proportional to $U_{l,\al}$ in
 eqn.~(\ref{eqn1}). It is important that this self-consistency does not
 refer only to global properties, but actually it refers to the
  local values of all charge- and spin densities.

  At the end, we comment on the parameter $\Delta_{l}$ in eqn.~(\ref{eqn1}):
   If we want to
  generate a globally ferromagnetic state, we start everywhere with
   positive values $\Delta_l =1$ eV, which are then switched off
   after the first iteration. In contrast, for the generation of spin
   glass states we use a more complicated procedure: At first, we start
   with a $\Delta_l$, which is $\ne 0$ only for one Fe site (for which
   we usually take one surrounded by a large number of Ni atoms, since
   there seems to be a rather strong and positive exchange interaction
   between Fe and Ni). Then, in  five iterations of
   eqn.~(\ref{eqn1}), a nontrivial  spin configuration develops
       in the
   whole sample,  which is already almost
    self-consistent concerning the final moment directions, but not concerning
    the amplitudes; i.e.~the spin configuration obtained
    has different signs of the spin polarization
   at different atoms. This distribution of signs may look random:
   Actually, it is not, since it reflects correlations due to
 the interactions in the system. So we now continue in iteration
   number 6 with transient $\Delta_{l}$ values of magnitude 1 eV everywhere,
   but with different signs at different  $l$,
    namely the signs of the local spin polarization determined
    in the first five iterations. For the following iterations, the
$\Delta_{l}$
 are then switched off completely, and the equations are iterated to
self-consistency, which can take 100 iterations or even more. In this way,
our spin glass configurations are generated, %  which have already
%been mentioned above and are to be discussed below, and
 which still
have self-consistently determined magnetic short-range order.
\section{Results}
\subsection{Magnetic configuration}

 In Fig.~1, we present the magnetic phase diagram as determined
 ex\-peri\-men\-tally by \cite{Boettger,HesseBoett}.
 Here, for 0.04 $\le y \le 0.14$, one
 observes re-entrant spin glass behaviour, i.e.~a transition from
 paramagnetic via ferromagnetic to spin glass behaviour, if the
 temperature is decreased. However, already at
 this point one should note that there exists a particular region between the
 ferromagnetic and the spin-glass regions, where both states apparently
 co-exist as metastable states. In fact, in Fig.~2 we show the results
 of our numerical simulation with ferromagnetic preparation:
 Here, up to $y\approx 0.07$, a nice
 agreement  between the experimentally determined average moment and our
 results is found, whereas for larger $y$ our calculated behaviour
 apparently continues as before in the ferromagnetic state with
 gradually decreasing average, whereas experimentally, there is a more
 drastic decrease to the spin glass phase. However, since in our
 calculation we use periodic boundary conditions with a periodicity
 length of only three fcc lattice constants and in view of the other
 approximations of our method, this discrepancy is perhaps not
 astonishing. Namely, throughout the same
 concentration region, by the above-mentioned spin glass preparation, we can
also
 get the already mentioned
 metastable spin glass states with essentially vanishing average
 moment. A typical plot of the moment distribution in such a spin glass
 state is presented in Fig.~3 for a disordered Fe$_{0.65}$Ni$_{0.35}$ system.

  Comparing the energy of the ferromagnetic and the spin glass states,
  we find that our
 ferromagnetic states are somewhat more favourable by energies of $\approx
 6$ meV per atom. However, as mentioned above, the accuracy of our
 method is not reliable enough to make this a significant
 statement.
 %On the other hand, the temperature scale, $T\approx 60 $ K,
 %corresponding to the energy difference, fits nicely to Fig.~1.
 In fact, experimentally, see Fig. 1, the spin glass state seems to have lower
 energy; but at least the energy difference, corresponding to $T \approx 60K$,
is
 of the same order.

 In Fig.~4, we present histograms for the magnetic moment distributions of
  Fe-, Ni-, and Mn-atoms for 11 different Mn concentrations $y$ of
  the {\it ferromagnetic states} calculated by our method for
  \linebreak
  (Fe$_{0.65}$Ni$_{0.35}$)$_{1-y}$Mn$_y$. Fig.~5 exhibits the average moments
  versus concentration.
  Interestingly, the Ni moments
  have apparently always well defined moments around $0.5$ $\mu_B$, whereas the
Fe
  moment distributions are rather broad, e.g.~they are spread between
  $\sim 0.5$ and $\sim 2.5$ $\mu_B$ , for $y=0.04$. Of course, details
  of the Fe distribution depend on the Mn concentration $y$, e.g.~for
  $y=0$ only Fe moments above 1.5 $\mu_B$ are observed, whereas for
  $y\approx 0.05$ also moments around 0.5 $\mu_B$ appear, but generally,
  the $y$-dependence of the $Fe$-moment distributions does not look
  drastic.

  With the Mn moments itself, this is different:  Some of the Mn
  atoms are polarized parallel to the majority spin direction, but the
  majority of the Mn atoms seems to be polarized antiparallel, as one can
  see from Fig.~5. Moreover, the magnitude of the Mn moments can vary
 drastically from site to site, which is not astonishing, since this was
already true for $\al$-Mn, \cite{Suess}.

For the {\it spin glass configurations} it is characteristic, see Fig.~6, that
 the Ni moments are no longer spread around 0.5 $\mu_B$, but
around zero. Further, for $y=0.2$ both the Fe and the Mn moments are
almost homogeneously distributed between $\approx -2$ $\mu_B$ and
 $\approx +2$ $\mu_B$. (For \hbox{$ y$ =0}, the Fe moments are distributed
 between $-2.5$ and $+2.5$ $\mu_B$.)

In Fig. 7-9 we compare our calculated distributions for the Fe mo\-ments with
hyperfine field measurements obtained by M\"ossbauer spec\-tro\-sco\-py
experiments \cite{HuckHess}.
For y=0 and y=0.039 (Fig. 7 and 8) there is good agreement between the
experimental hyperfine field distribution and the magnetic moment distribution
in the
ferromagnetic state. For higher Mn-concentrations, e.g. y=0.102, the spin glass
state corresponds better to the experimental result (Fig. 9).
For each of this three con\-cen\-tra\-tions we assume the same conversion
factor of
$14.5$ T per $\mu_B$.

According to \cite{HuckHess} the oscillatory structure of the hyperfine field
dis\-tri\-bu\-tion is
of no physical relevance, but is simply a consequence of the fit procedure used
by the authors.

A natural question is, whether the magnetic moments depend
on the number of neighbouring atoms of the same or different
kind. Therefore, in Fig.~10, we present some kind of plot, which allows
answers on this question: E.g., from this figure it is obvious that high
Fe moments are favoured by a high number of Ni neighbours, and
disfavoured by a high number of Mn neighbours, whereas the Ni moments
themselves are rather insensitive on the neighbourhood. However, the
 spread of the distributions in Fig.~10 around the average data points
 should be kept in mind, when stating these tendencies, which can of
 course be quantified only on average, see the open circles in Fig.~10.
  Finally, it should be noted again that the magnitude of the Mn moments
  range between 0 and $\approx 2$ $\mu_B$, with typical values around 1
  $\mu_B$. Fig.11 presents similar plots for the spin glass
  configurations, where the width of the distributions is impressive,
  and the tendencies concerning Ni neighbours in case of Fe atoms are
  similar.

  We have tried similar plots as in Fig.~10 and Fig.~11 also for the 2nd-
   and 3rd-nearest neighbour shell, however with non-conclusive results.
   Therefore, those plots are omitted.
 \subsection{Simplified spin model}
 In the following we try to describe the results for the magnetic behaviour
 with a simple molecular-field type {\it ansatz}. For the determination
 of the %average
 magnetic moments $\mu_l^z $ at the
 different sites $l$ of the sample we have tried the equation
 \beqn \label{Ansatz}
%\langle
\mu_l^z
%\rangle_l
 = \sum_{k=1}^{42} c_{A(l),A(k)}
% \langle
 \,\,\cdot\mu_k^z %\rangle_l\,.
 \eeqn
%Here the brackets $\langle ...\rangle_l$ mean an average over the sites $l$,
 In eqn.~(\ref{Ansatz}), the sum is over the 42 neighbours
 of site $l$ up to the third-nearest neighbour shell, and
 there are nine coefficients per shell, namely
 $c_{A(l),A(k)} \in \{ c_{Fe,Fe}; c_{Fe,Ni};
 c_{Fe,Mn};...;
 c_{Mn,Fe}; c_{Mn,Ni}; c_{Mn,Mn} \}$, depending on the occupation
  of the sites $l$ and $k$ by the respective kind of atoms. Qualitatively,
the terms  $c_{A(l),A(k)}\cdot \mu_k$ can be understood as composed of a local
 susceptibility $\chi_l$ times a local effective field $h_{l,k}\propto
\mu_k^z$.
In fact, fitting the % nine
coefficients  % per shell
% by the NAG procedure E04FDF
to our results for $\mu_l^z$, which we obtained
by our itinerant approach, with the ansatz of eqn.~(\ref{Ansatz}),
 we obtain the results presented in Fig.~12. Although these results
 depend on the Mn concentration $y$, and although the statistical error
 bars, which  we have omitted to show, are rather large, namely typically
  $\pm 0.06$, some conclusions can be drawn from Fig.~12:
 In the ferromagnetic state, the magnetization of Fe atoms is
 strongly influenced
 by the Ni neighbors in the first neighbour
  shell, and up to $y\approx 0.1$ also by those of the second neighbour shell.
 In fact, while  $c_{Fe,Ni}$ is as large as 0.3 for $y$=0 and increases
 up to $0.4$ for larger $y$, $c_{Fe,Fe}$ is only around 0.06 for $y$=0 and
  even as low as 0.025 for $y > 0.1$; however, it
 should of course be taken into account that $\mu_{Ni} \approx 0.5$, while
 $\mu_{Fe}$ can be three or five times as large, so Ni is dominating, but
 not strongly. Moreover, the
  coefficient $c_{Fe,Mn}$ is negative over a considerable $y$-range and
particularly strong in the 2nd-neighbour shell for small $y$-values,
  similar as $c_{Mn,Mn}$, %and $c_{Mn,Fe}$,
 whereas $c_{Fe,Fe}$ is
positive in the ferromagnetic state, as mentioned above, contrary
 to a formerly wide-spread assumption (\cite{Kloss}).  Also interesting
 is the strong scatter of the coefficients for Mn, i.e.~the third column.
 This might show that the Mn moments tend to get ''frustrated'', however a
 definite statement cannot be made in view of the small number of Mn atoms.

  Most important, however, is the fact
   that for the spin glass states (for which we omit the corresponding
   plots \cite{Paintner}) the coefficients are different from those of the
ferromagnetic
   case presented in Fig.~12: The main difference concerns $c_{Fe,Ni}$,
    which becomes as high as  2.3 for $y$=0, 1.5 for
   $y$=0.1 and  $1.4$ for $y=0.2$. This is understandable, since
   in the spin glass state also $\mu_{Ni}$ decreases in magnitude from
   $\approx 0.5$ $\mu_B$ to much smaller values (see above); however due to
    the increase of $c_{Fe,Ni}$, the Ni moments remain important for the
    Fe moments.
     A second important difference between the ferromagnetic
     and the spin glass states is that
   $c_{Fe,Fe}$, which was positive for ferromagnetic preparation, is
   negative ($\approx -0.1$) for spin glass preparation.
  So the implicit assumption made
   with the above-mentioned {\it ansatz}, namely that the $c_{A(i),A(k)}$
coefficients
   should not depend on the global magnetic state, is obviously  not true.
   This means, one should better stay with the itinerant
   description, realizing that here the spin configuration is not fixed
   by given interaction coefficients as in  Heisenberg spin glasses:
   Instead, the effective interactions depend on the
   global itinerant state and are different for a ferromagnetic
   and an itinerant spin-glass configuration of the system. Thus the
itinerant description is really necessary.
  \subsection{Electrical resistivity in the magnetic state}
It is perhaps not widely known that in a disordered crystalline or amorphous
metallic system not only  single particle properties, as just discussed,
 but even
two-particle properties as the electrical resistivity can be calculated
by our semi-empirical LCAO formalism, namely by means of the Kubo
formalism, \cite{KreyWid}. In fact, in \cite{KreyWid}, the
Kubo formalism has been successfully applied for the numerical
calculation of the resistivity in the magnetic state of amorphous Fe/Zr
alloys, and there we have also described how the conductivity
$\sig_{x,x}$ at T=0 K as a function of the Fermi-energy $E_f$ can be
calculated from the formula

\beqn \label{eqn4}
\sig_{x,x}(E_f) =\frac{e^2\hbar\eta}{\Omega\Delta E}
\sum_{\eps^{(\nu )}\in \Delta E}
\langle v_x \vec{c}^{\,(\nu )}|[(E_f-H)^2+\eta^2]^{-1}v_x \vec{c}^{\,(\nu
)}\rangle \, .
\eeqn

Here the sum is over all $\nu$ with $\eps^{(\nu)}
\in\Delta E$; further, $\Omega$ is the volume per atom, $\vec{c}
^{\,(\nu)}$ an eigenstate of
eqn.~(\ref{eqn1}) with energy $\eps^{(\nu)}$ in the interval $\Delta E$
around $E_f$, and $\eta$ is an effective line broadening parameter, which
should be of the order of 1/2 the typical energy distance between
neighbouring eigenvalues. Of course, the results depend on the choice of
$\Delta E$ and of $\eta$, and additionally one has to perform sample averages
and averages over the directions $x$, $y$, and $z$ of the conductivity.
Then, as we see in the Fig.~13 below, and as we have already found in
 \cite{KreyWid}, the result is essentially independent on
$\Delta E$ and $\eta$\,\,\,:

In Fig.~13, we compare our results with the experiments in
\cite{BarnardBoett} and find good agreement, concerning the
dependence on the Mn concentration $y$, to the accuracy
considered, over the whole parameter range. However, there is one
discrepancy, which we do not yet understand at present, namely our
resistivities are always too large  by a \hbox{factor $\approx$ 5}, compared
with the
experiment. But
in view of the approximations involved in the evaluation of eq. (\ref{eqn4})
and in view of the large scale of resistivities of different metals this
discrepancy should perhaps not be taken too serious. In fact with
\hbox{$\eta \approx$ 0.02 eV}
we induce effectively inelastic scattering events corresponding to
\hbox{T $\approx$ 200 K}, and the effect of
\hbox{$\Delta E \approx$ 0.2 eV} may even be larger, whereas the
experimental values in our comparison are at 4 K. However, at 300 K,
according to \cite{BarnardBoett}, the experimental values would only
be enhanced with respect to those at 4 K by factors between 1.8 (for
$y$=0) and 1.1 for $y=0.084$. Additionally, the conductivity
should be influenced by transverse components of magnetic
moments, which we have neglected.

\section{Discussion}

We have studied the magnetic behaviour of Fe-Ni-Mn alloys \linebreak in the
sensitive
 regime around the Fe-Ni Invar-region, \cite{Wassermann}, i.e. \linebreak
 (Fe$_{0.65}$Ni$_{0.35}$)$_{1-y}$ Mn$_y$, where with increasing Mn
 concentration $y$ a transition from ferromagnetic to spin glass
 behaviour has been observed experimentally, \cite{Boettger,HesseBoett}.

Actually, with our semi-empirical LCAO approach we have found that both
ferromagnetic states and also itinerant spin glass states can be
prepared and are metastable in the whole concentration region
considered. These different states have roughly the same energy, i.e.~according
to our calculation, the ferromagnetic state appears to be slightly favoured
by $\sim 6$ meV per atom; however our method is not accurate enough for a
 definite statement.  In any case this implies that the itinerant magnetism of
these alloys is quite complicated and cannot simply be described by the
Heisenberg model, since one has e.g.~a broad distribution of different
values of the magnitudes of the local atomic moments, both for Fe and Mn.

Starting our iterations with a homogeneous ferromagnetic state, we
obtained good agreement with experimental results in the  ferromagnetic
state up to a Mn concentration of $y \approx 0.07$. For higher
concentrations, more realistic results (e.g.~the already mentioned
spin-glass states with
vanishing total magnetic moment, but non-trivial local moments) have
been obtained  systematically from a self-consistently prepared
 inhomogeneous
state with correct short range magnetic order. Thus there seems to be some
kind of preparation dependence of the equilibrium state. After all, this is
typical for a situation with possible spin-glass behaviour and not
in contradiction to the experimental situation, see \cite{Boettger}.

Finally, we have also calculated the electrical resistivity in the
magnetic state in our LCAO formalism by an implementation of the Kubo
formalism (see \cite{KreyWid}), and we obtained good agreement with
experiments apart from the fact that our conductivities are generally
too low by a factor of $\approx$ 5, compared with experiments.
%Although at
%present we can only speculate on the reason of this discrepancy, it is
%natural to guess that it may have to do with the fact that throughout
%our calculation the transverse spin components have been neglected.
The problem, whether the conductivity in the magnetic state will
%perhaps be enhanced by the additional scattering channels opening
change with the
  introduction of transverse spin components, remains to
   be studied in the future.

\subsection*{Acknowledgements}
Two of the authors (U. K. and F. S.) would like to thank the Institute of
Molecular Physics
of the Polish Academy of Sciences in Pozna\'n for its hospitality and
particularly Dr. S. Krompiewski and Prof. J. Morkowski for useful discussions.

\newpage
\parindent-3em
{\bf Figure Captions}

Fig. 1:
Magnetic phase diagram for the system
$(Fe_{0.65}Ni_{0.35})_{1-y}Mn_y$, from [6,8].

Fig. 2: Magnetization of the alloy system
$(Fe_{0.65}Ni_{0.35})_{1-y}Mn_y$ versus Mn-concentration $y$.

Fig. 3:
Typical moment distribution of a spin glass state for a disordered
Fe$_{0.65}$Ni$_{0.35}$ alloy. The Ni moments are scaled
by a factor 10. The symbols $\oplus$ above and $\ominus$ below an atomic
position show
the local spin direction of the atoms at the intermediary state in the 6-th
iteration (see text).

Fig. 4:
Calculated histograms for the magnetic moment distribution of Fe-, Ni- and
Mn-atoms for
Mn concentrations y in the range $0<y<0.231$ (ferromagnetic preparation).

Fig. 5:
Average magnetic moments for the three different alloy components Fe,
Ni and Mn.

Fig. 6:
Calculated histograms for the magnetic moment distribution of Fe-, Ni- and
Mn-atoms for
Mn concentration $y\in\{0; 0.102; 0.204\}$ (spin glass preparation).

Fig. 7:
Calculated magnetic-moment probability histogram
 and measured hyperfine field probability distribution (dashed-line
fit) of the system $Fe_{0.65}Ni_{0.35}$. The simulation was done with
ferromagnetic preparation.  A conversion factor of $H=14.5$ $T$ per
$\mu_B$ is assumed.

Fig. 8: Calculated magnetic-moment probability histogram (ferromagnetic
preparation) of the system $(Fe_{0.65}Ni_{0.35})_{0.963}Mn_{0.037}$
and measured hyperfine field probability distribution (dashed-line
fit) of the system $(Fe_{0.65}Ni_{0.35})_{0.961}Mn_{0.039}$.

Fig. 9: Calculated magnetic-moment probability histogram (spin glass
preparation) of the system $(Fe_{0.65}Ni_{0.35})_{0.898}Mn_{0.102}$
and measured hyperfine field probability distribution (dashed-line
fit) of the system \\
$(Fe_{0.65}Ni_{0.35})_{0.919}Mn_{0.084}$.

Fig. 10:
Influence of the local neighbourhood on the magnitude of the magnetic
moments for final states with ferromagnetic preparation. In
the left three columns the magnitude of the Fe moments is plotted versus the
next neighbour number of Fe-, Ni- and Mn-neighbours for six different Mn
concentrations. In the middle three columns the same
is plotted for Ni, and in the right three columns for Mn.
The dashes represent the single values and the circles are the averaged values.

Fig. 11:
Influence of the local neighbourhood on the magnitude of the magnetic
moments for final states with spin glass preparation.

Fig. 12: The coefficients $c_{A(l),A(k)}$ from equation (4) are
plotted against Mn concentration y for states with ferromagnetic
preparation. The coefficients of the first row are related to next
neighbours. Those of the second and third row are related to second
resp.~third neighbour shells.

Fig. 13:
Measured and calculated conductivities against Mn concentration for several
combinations of the parameters $\eta$ and $\Delta E$.

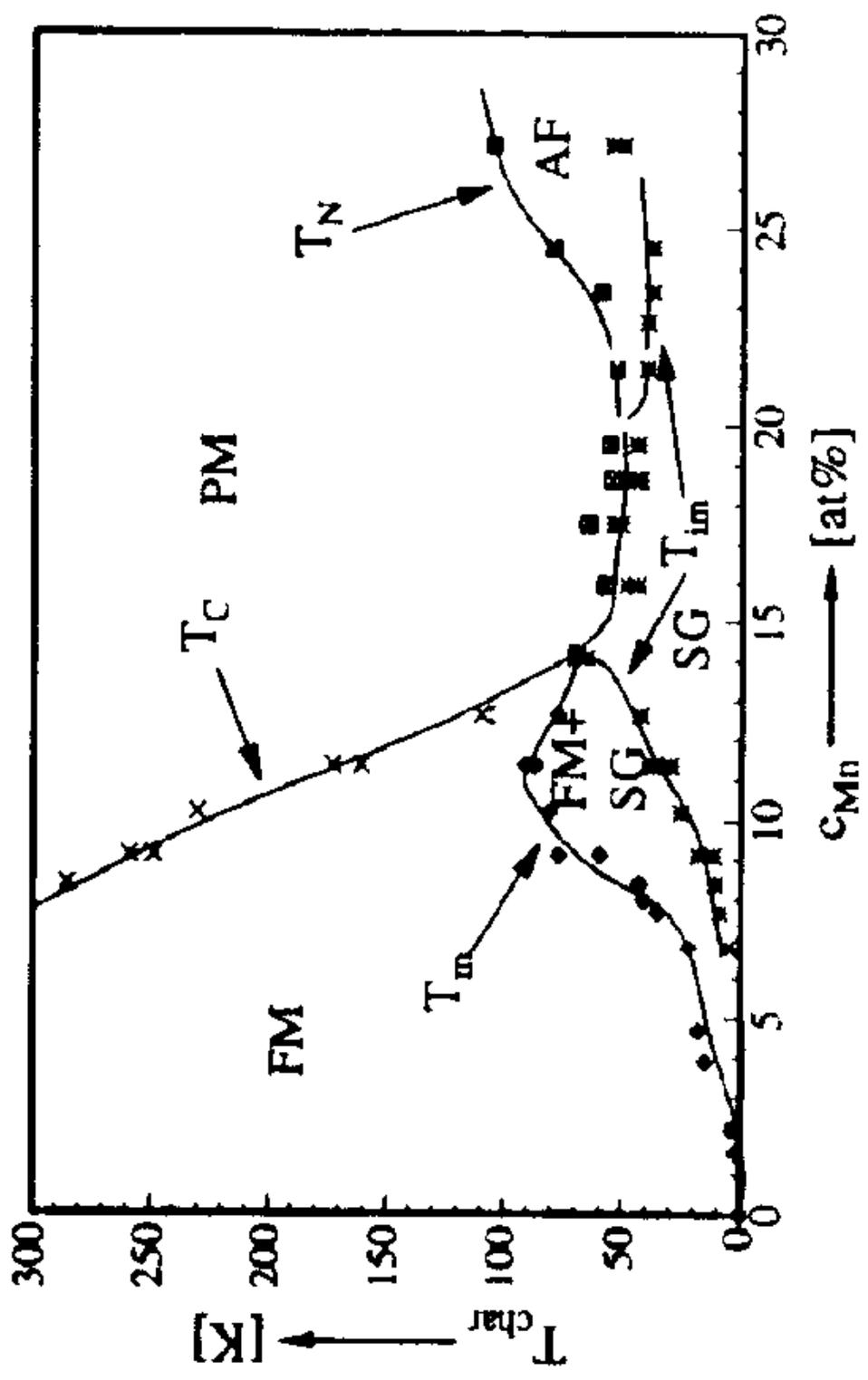
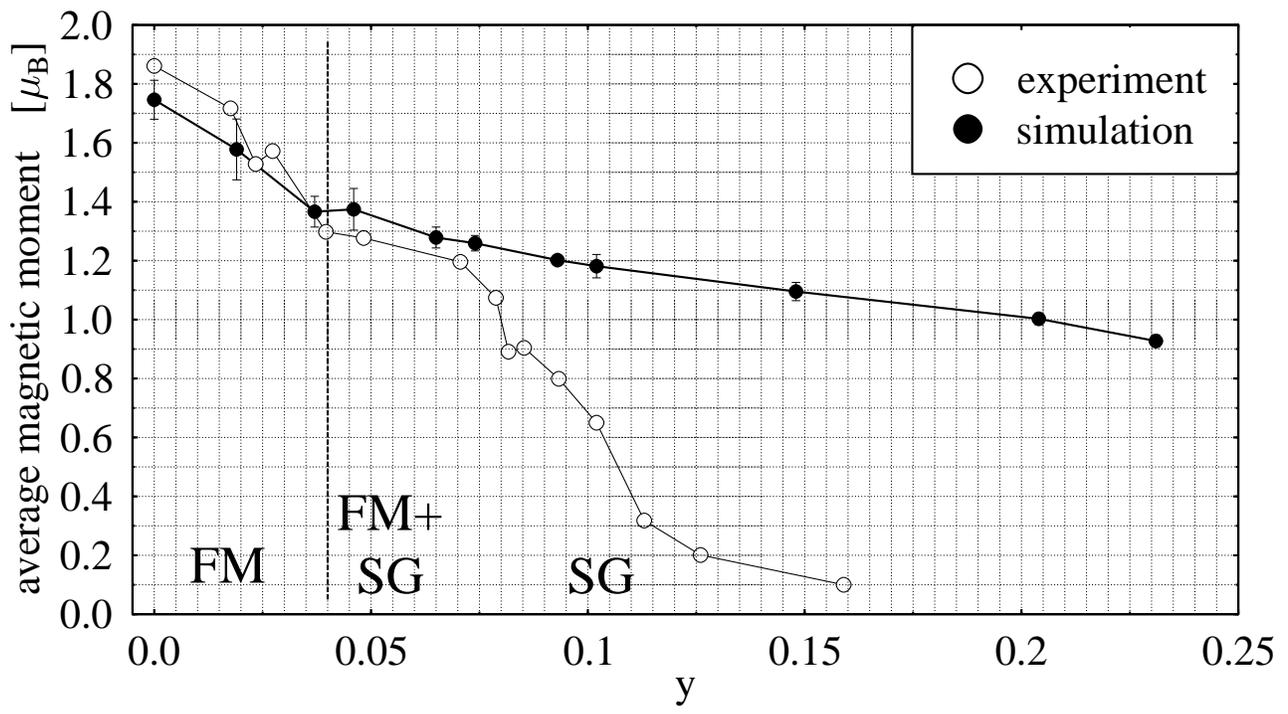
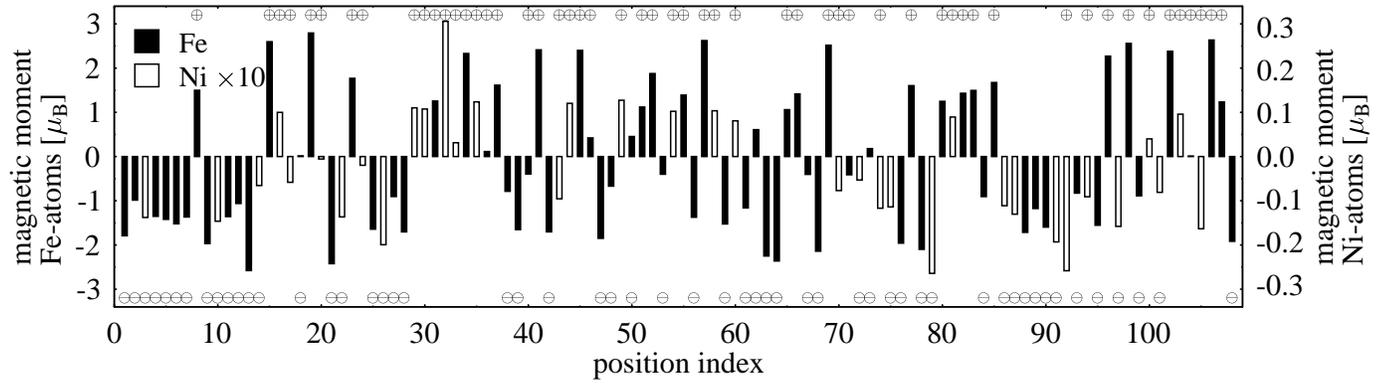
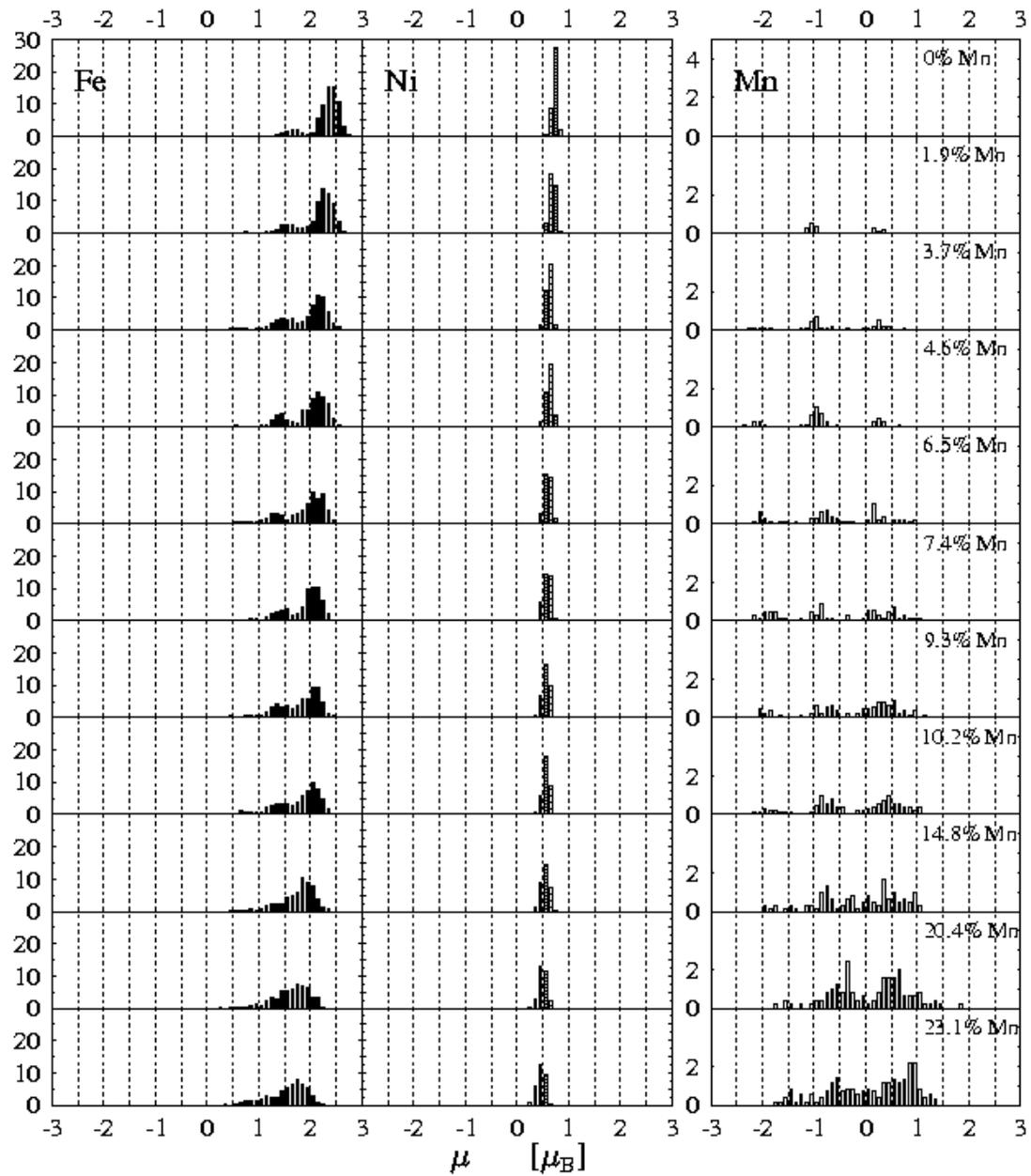
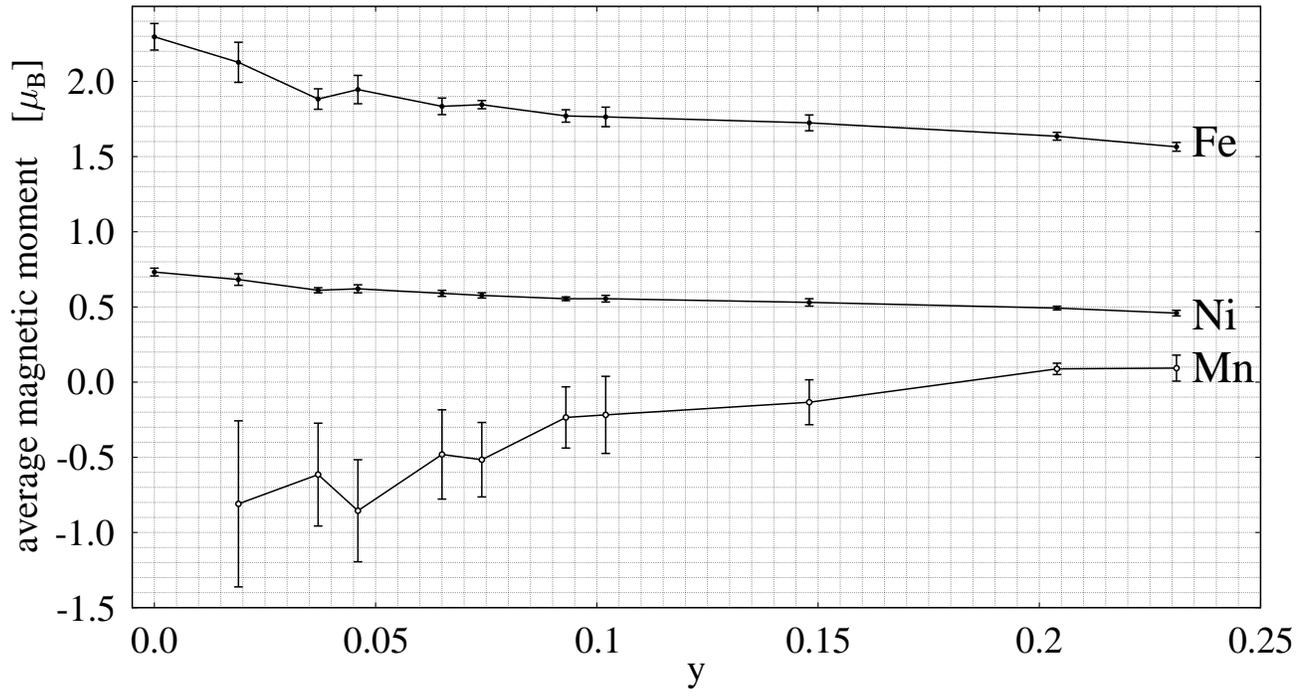
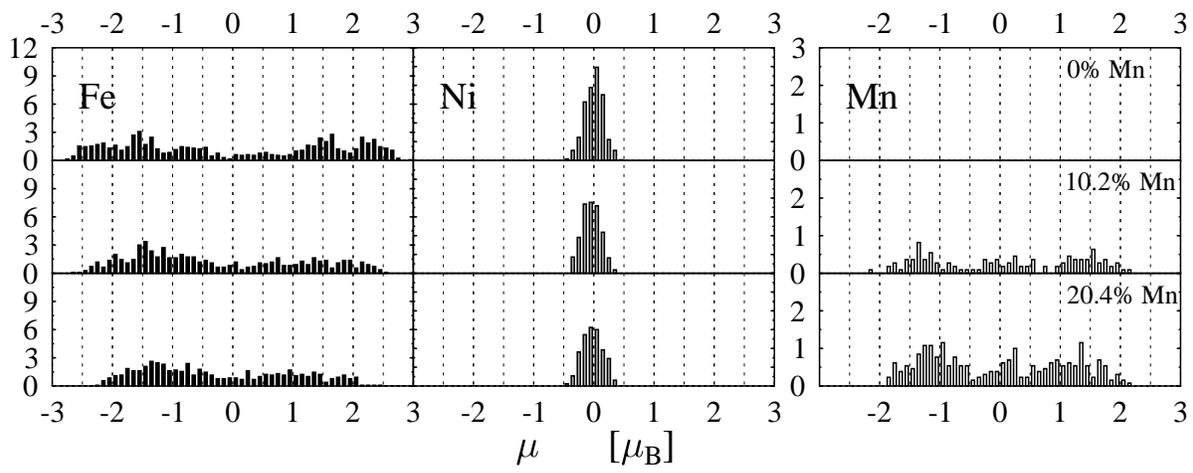
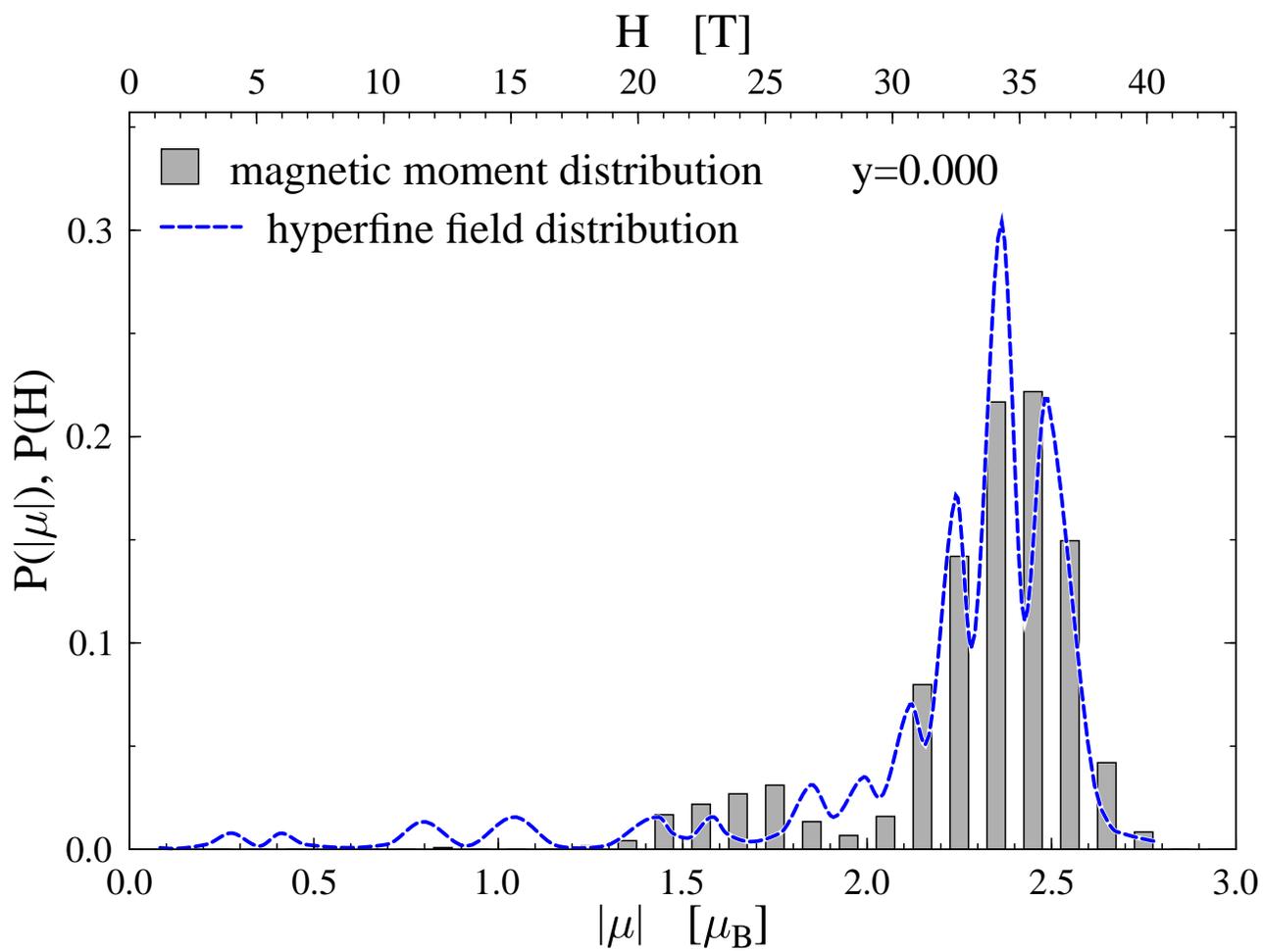
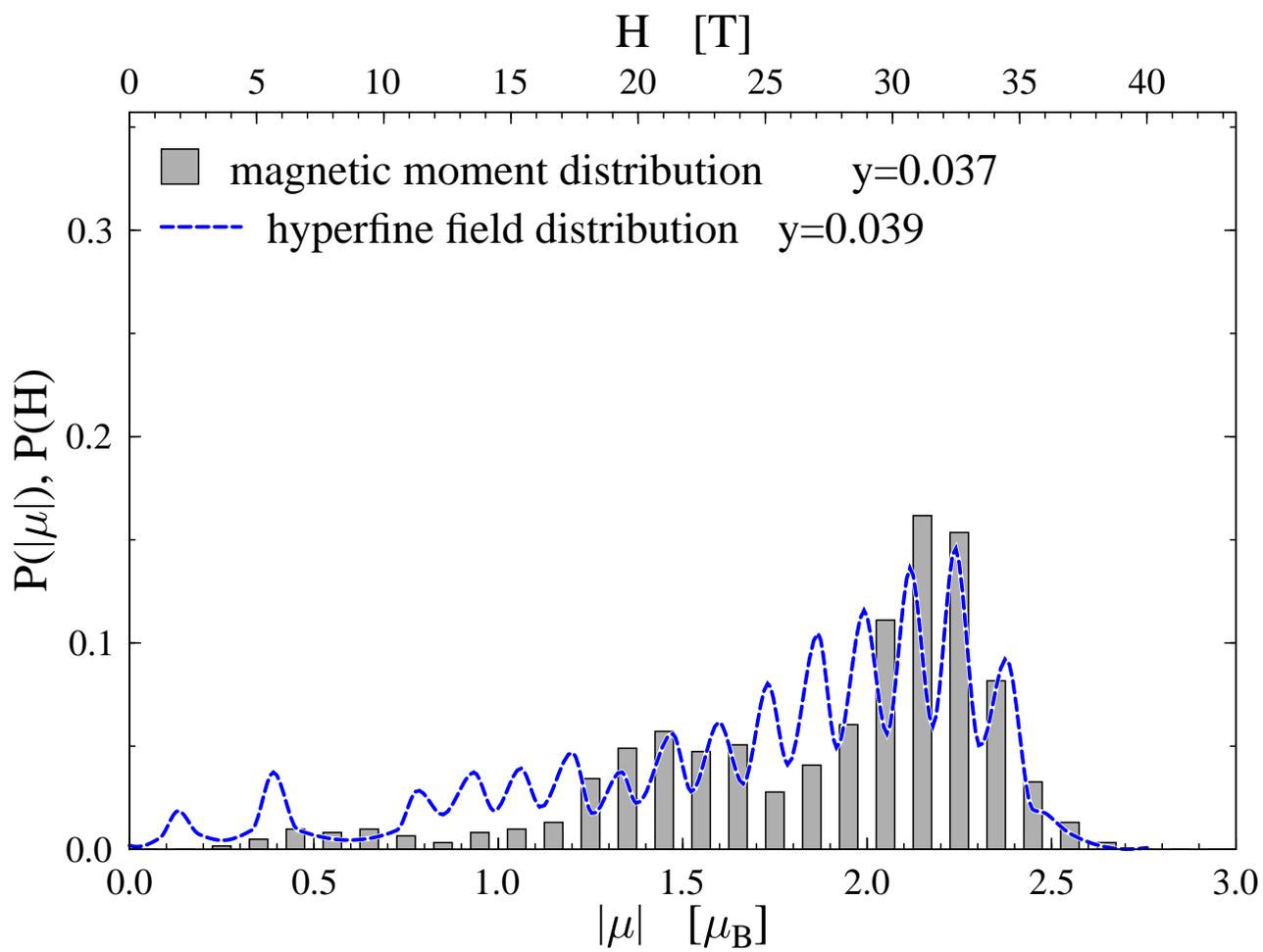
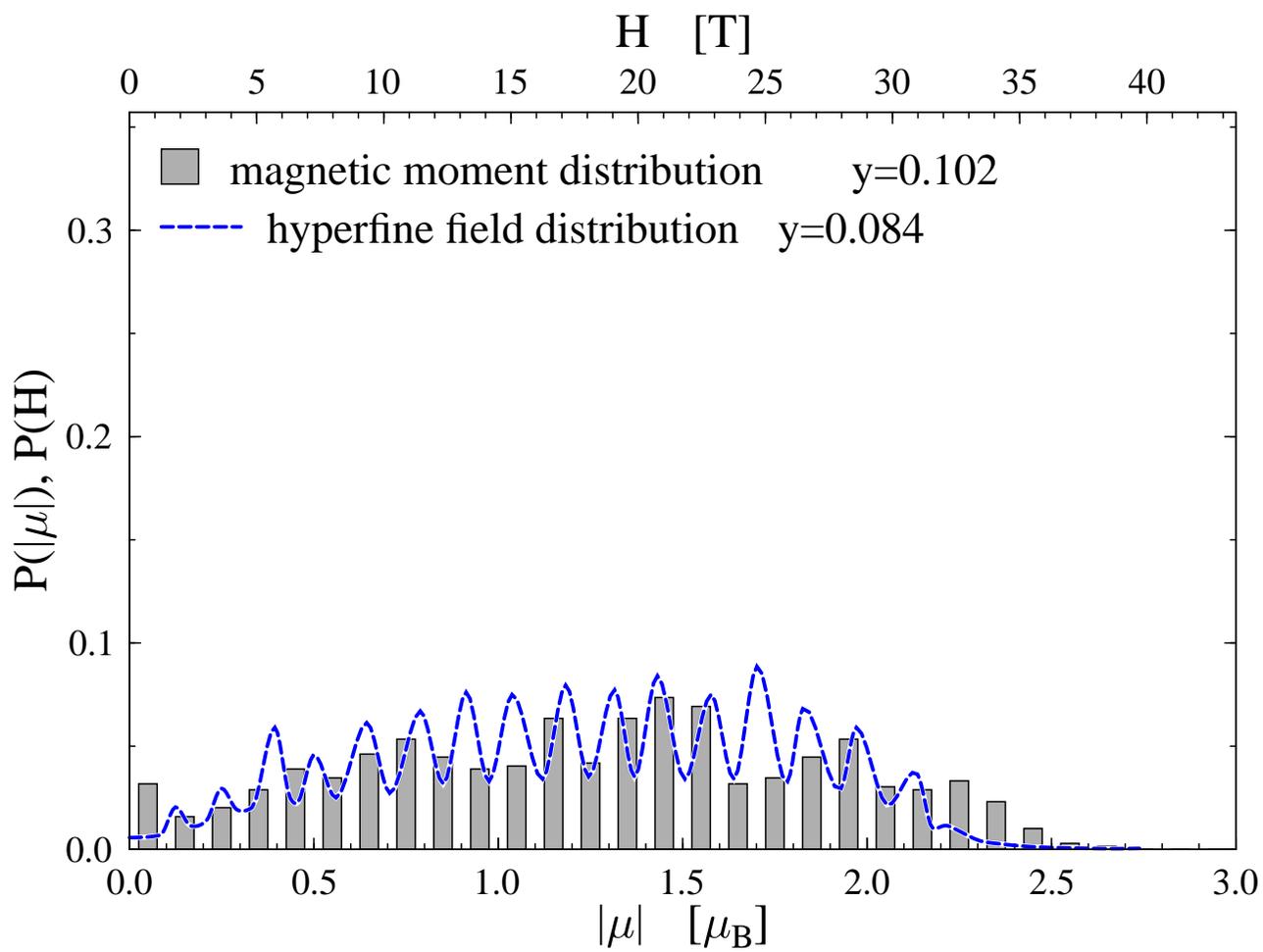
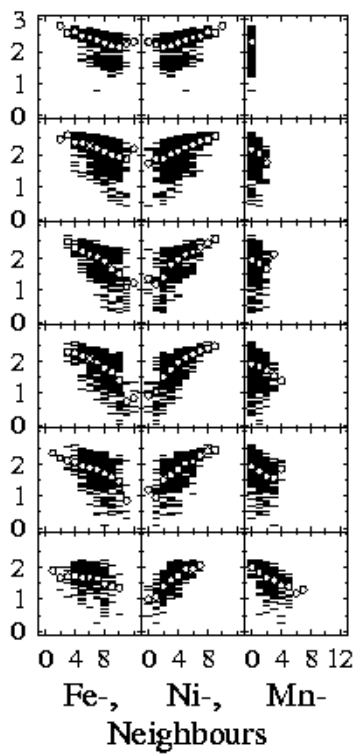
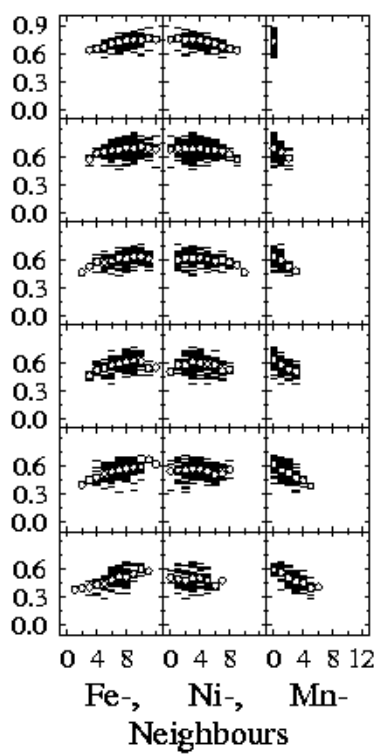
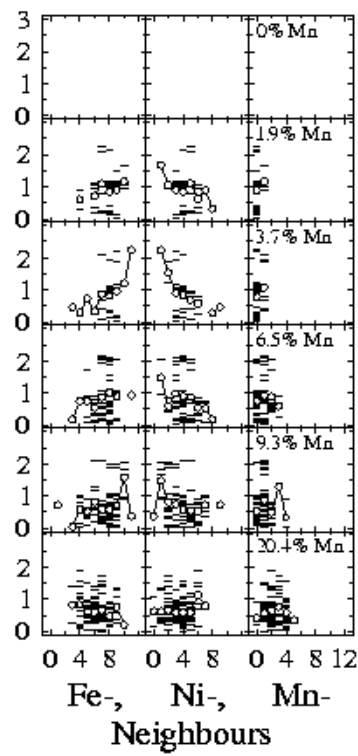
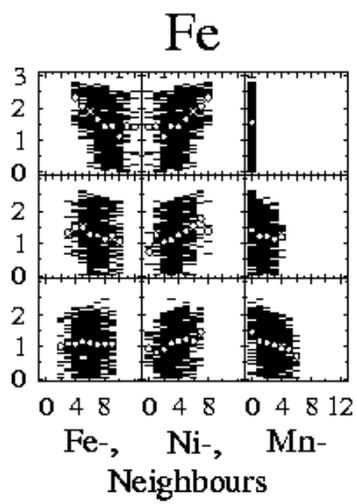
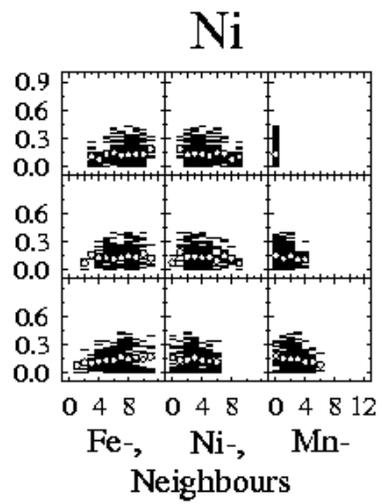
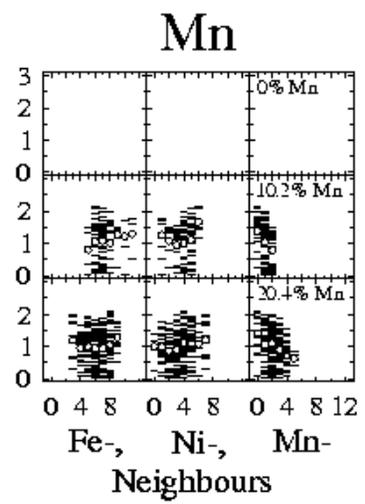
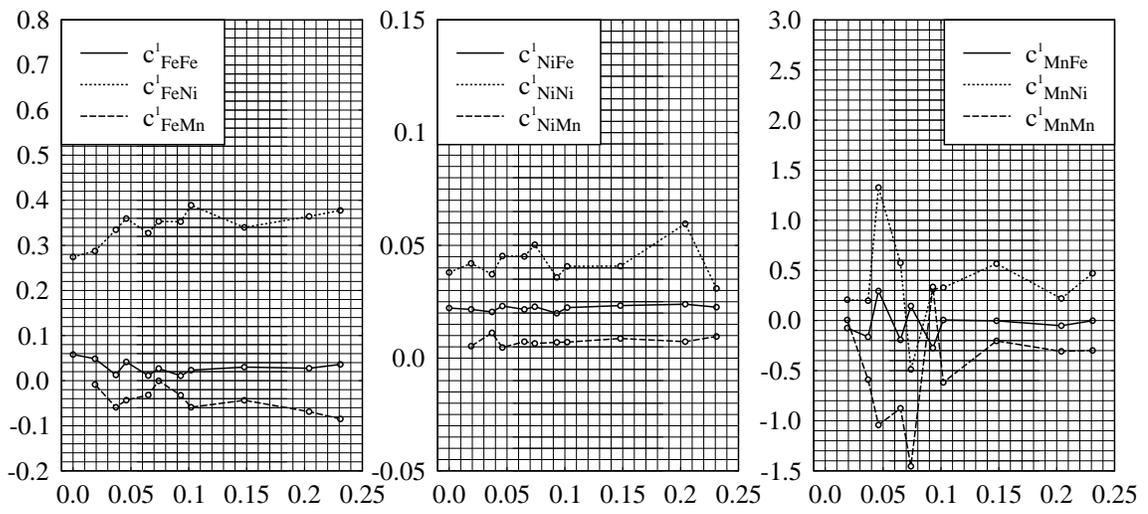
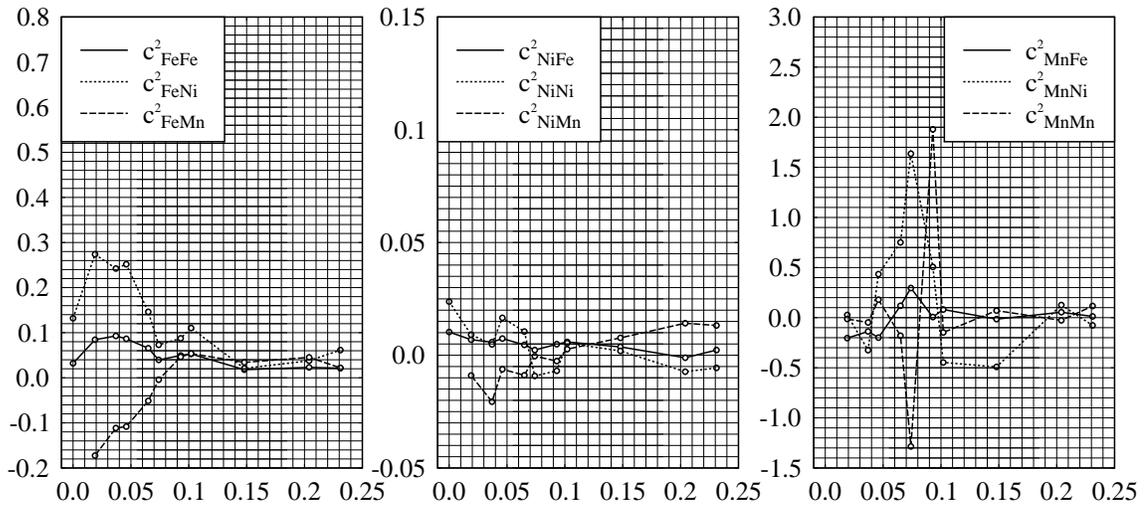
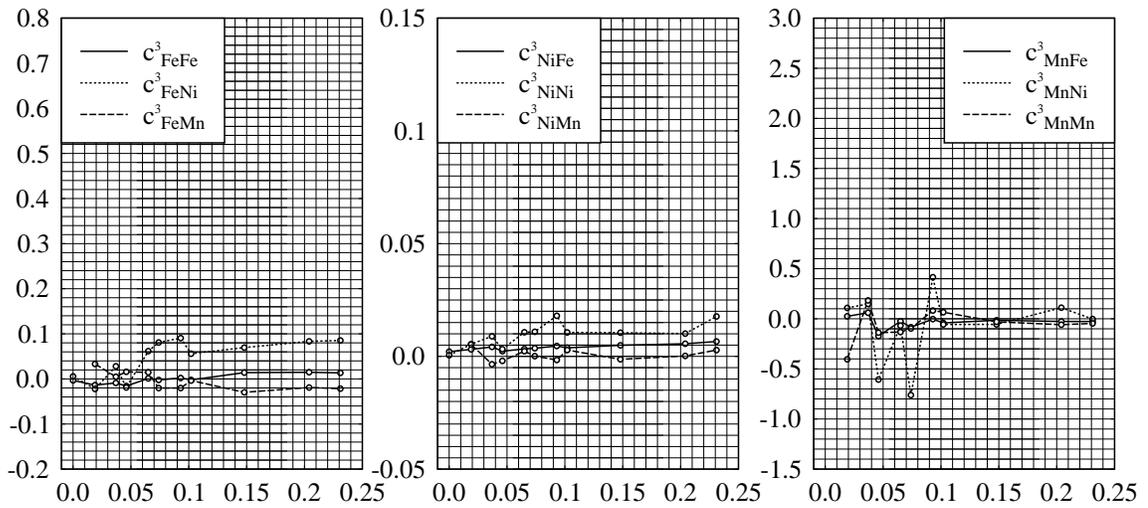
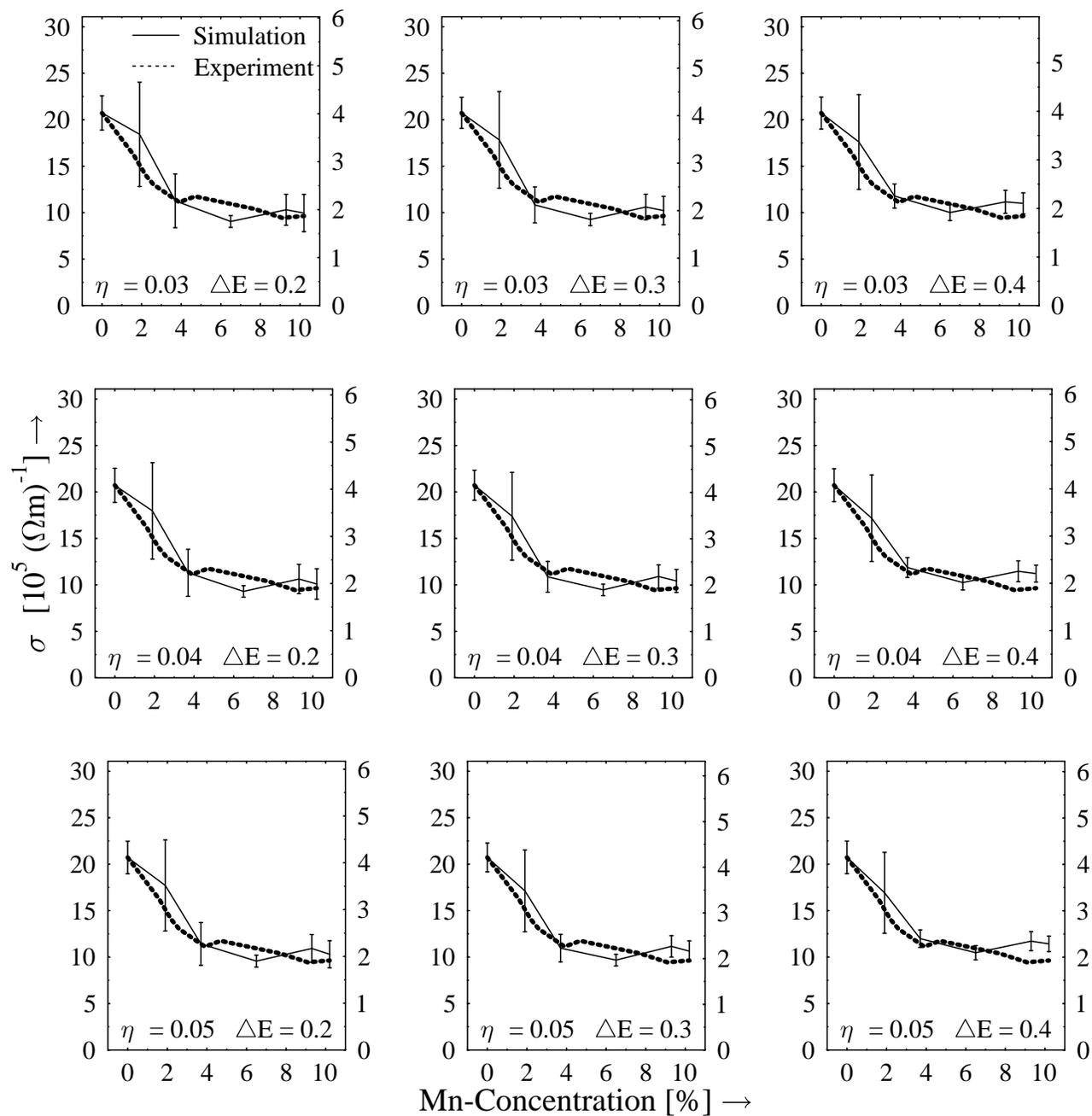
\end{document}